\documentclass[11pt]{article}
\usepackage{latexsym}

\usepackage{amssymb}
\usepackage{amsmath}

\numberwithin{equation}{section}

\newcommand{\be}{\begin{equation}}
\newcommand{\ee}{\end{equation}}
\newcommand{\ba}{\begin{eqnarray}}
\newcommand{\ea}{\end{eqnarray}}
\newcommand{\baa}{\begin{array}}
\newcommand{\eaa}{\end{array}}
\newcommand{\nn}{\nonumber \\}
\newcommand{\nr}[1]{(\ref{#1})}

\newcommand{\fr}[2]{{\frac{#1}{#2}\,}}
\newcommand{\fra}[2]{\textstyle{\frac{#1}{#2}\,}}  %%Smaller size!
\newcommand{\mn}{{\mu\nu}}

\newcommand{\half}{{\textstyle{1\over2}}}

\newcommand{\rapidity}{{y}}

\def\Tr{{\rm Tr\,}}

\def\CL{{\cal L}}

\def\gsim{\raise0.3ex\hbox{$>$\kern-0.75em\raise-1.1ex\hbox{$\sim$}}}
\def\lsim{\raise0.3ex\hbox{$<$\kern-0.75em\raise-1.1ex\hbox{$\sim$}}}

\begin{document}

\begin{titlepage}
\begin{flushright}
HIP-2008-31/TH\\
\end{flushright}
\begin{centering}
\vfill

{\Large{\bf Casimir energy in the gauge/gravity description\\[1ex] 
of Bjorken flow?}}

\vspace{0.8cm}

\renewcommand{\thefootnote}{\fnsymbol{footnote}}
% changes the title page footnote counter style to \fnsymbol

K. Kajantie$^{\rm a,b}$\footnote{keijo.kajantie@helsinki.fi},
Jorma Louko$^{\rm c}$\footnote{jorma.louko@nottingham.ac.uk},
T. Tahkokallio$^{\rm a,b,d}$\footnote{ttahko@uvic.ca}

\setcounter{footnote}{0}
% resets the footnote counter after the title page

\vspace{0.8cm}

{\em $^{\rm a}$%
Department of Physics, P.O.Box 64, FIN-00014 University of Helsinki,
Finland\\}
{\em $^{\rm b}$%
Helsinki Institute of Physics, P.O.Box 64, FIN-00014 University of
Helsinki, Finland\\}
{\em $^{\rm c}$%
School of Mathematical Sciences, University of Nottingham,
Nottingham NG7 2RD, UK\\}
{\em $^{\rm d}$%
Department of Physics and Astronomy, University of Victoria, 
Victoria, British Columbia, V8W 3P6 Canada\\}

\vspace*{0.8cm}

\end{centering}

\noindent
In the AdS/CFT description of four-dimensional QCD matter undergoing
Bjorken expansion, does the holographic energy-momentum tensor contain
a Casimir-type contribution that should not be attributed to thermal
matter?  When the bulk isometry ansatz that yielded such a Casimir
term for $(1+1)$-dimensional boundary matter is generalised to a
four-dimensional boundary, we show that a Casimir term does
not arise, owing to singularities in the five-dimensional bulk
solution.  The geometric reasons are traced to a difference between the
isometries of AdS${}_{3}$ and AdS${}_{d+1}$ for $d\ge3$. 

\vfill \noindent

\vspace*{1cm}

\noindent
Revised December 2008

\vspace*{1ex}

\noindent
Published in 
Phys.\ Rev.\ D {\bf 78}, 126012 (2008)

\vfill

\end{titlepage}

\section{Introduction}

Ultrarelativistic heavy ion collisions can be modelled by the
scale-free Bjorken expansion \cite{bj} 
in which the energy density $\epsilon$ and the temperature 
$T$ depend on the proper time $\tau$ as\footnote{We use the
Minkowski coordinates $(t,x,x^2,x^3)$, 
in which $ds^2=-dt^2+dx^2+{(dx^2)}^2+{(dx^3)}^2$, $x$ is the 
longitudinal spatial coordinate and 
$(x^2, x^3)$ are the transversal spatial coordinates. 
The collision is at $t=0=x$. 
In the quadrant to the future of the collision, $t > |x|$, 
we use the Milne-like coordinates $(\tau, \rapidity, x^2,x^3)$,
in which $\tau=\sqrt{t^2-x^2}$
with $0< \tau < \infty$, 
$\rapidity=(1/2)\log[(t+x)/(t-x)]$
with $-\infty < \rapidity < \infty$
and the metric reads $ds^2 = -d\tau^2+\tau^2d\rapidity^2+{(dx^2)}^2+{(dx^3)}^2$.}
\be
\epsilon(\tau) \sim T^4(\tau) \sim {1\over\tau^{4/3}},
\qquad T(\tau)\sim{1\over\tau^{1/3}}.
\label{epsbj}
\ee
If also the shear viscosity $\eta$ is scale free, 
$\eta\sim T^3$, relativistic hydrodynamics
predicts \cite{hk} that $\epsilon$ and $T$ are corrected by terms of order
$1/(\tau T)\sim 1/\tau^{2/3}$, 
\be
T(\tau)=T_f\left({\tau_f\over\tau}\right)^{1/3}-{c\over\tau}, 
\label{tvisc}
\ee
where 
$c$ is a positive constant and the 
normalisation of $T(\tau)$ at large $\tau$ is encoded in the 
constants $T_f$ and~$\tau_f$. 
The physical basis of this fluid model is that for large
nuclei moving in the $x$ direction
one may take the transverse size to be infinite so that there is no
dependence on the transverse coordinates. Also the incident energy
is taken to be infinite, so that there is no preferred longitudinal frame 
and the flow is a similarity flow, $v=x/t$, and $\epsilon(\tau)$ does not
depend on the rapidity coordinate~$\rapidity$. Realistic physical
conditions violate these assumptions in obvious ways; 
also, there is a hadronisation
transition at some $T=T_c$.

When QCD matter is approximated by a conformal fluid, 
AdS$_5$/CFT$_4$ duality provides a tool for predicting 
the thermodynamical 
and hydrodynamical properties of the matter. 
Conceptually, the approach is very simple: 
find the 
relevant solution of the bulk Einstein theory with 
a negative cosmological constant, and compute from it 
the boundary energy-momentum tensor, 
the temperature $T(\tau)$ and the entropy density~$s(\tau)$. 
What complicates the task is that the 
relevant bulk solution is known 
only in terms of asymptotic large $\tau$ expansions 
\cite{jp,jp2,nakamura1,nakamura2,janik1,janik2,%
kt,kovchegov,bbhj,buchel,Heller:2008mb,Kinoshita:2008dq}. 
One finds for 
$T(\tau)$ an expansion in $1/(\tau T)\sim 1/\tau^{2/3}$, 
generalising~\nr{tvisc}:
\be
T(\tau)=T_f\left({\tau_f\over\tau}\right)^{1/3}+{\eta_0\over\tau}
-{\eta_0^2(1-\log2)\over T_f\tau_f^{1/3}}
{1\over\tau^{5/3}}+{\eta_0^3A\over (T_f\tau_f^{1/3})^2\tau^{7/3}}
+{\eta_0^4B\over(T_f\tau_f^{1/3})^3\tau^3}+\cdots\, ,
\label{tconfexp}
\ee
where in the second term on the right-hand side we have 
\be
\eta_0=-{1\over6\pi} , 
\ee
following from the viscosity prediction $\eta/s=1/(4\pi)$, 
and the third term on the right-hand side 
contains the corresponding quantities from conformal second
order hydrodynamics \cite{brsss,Natsuume:2007ty,Natsuume:2007tz}. 
The unknown dimensionless 
constants $A$ and $B$ would correspond to third
and fourth order hydrodynamics. 
For the energy density we then have, using units in which 
$T_f\tau_f^{1/3}=\sqrt2/(3^{1/4}\pi)$ \cite{janik2,brsss}, 
\ba
\epsilon(\tau)&=&{3\pi^2\over8}N_c^2\left({(T_f\tau_f^{1/3})^4\over\tau^{4/3}}+
{4\eta_0(T_f\tau_f^{1/3})^3\over\tau^2}
+{2\eta_0^2(1+\log4)(T_f\tau_f^{1/3})^2\over\tau^{8/3}}\right.+\nn 
&&\left.+{4\eta_0^3T_f\tau_f^{1/3}(-2+\log8+A)\over\tau^{10/3}}+
{\eta_0^4(-5+6\log^22+12A+4B)\over\tau^4}+\cdots \right).
\label{epsconfexp}
\ea

A question that these expansions do not directly 
address, however, is whether the holographic energy-momentum 
tensor provided by the AdS/CFT correspondence should be 
attributed in its entirety to excitations of the boundary matter. 
The holographic energy-momentum 
tensor could conceivably contain also a Casimir-type vacuum energy term, 
corresponding to a QFT `vacuum' 
that is not the conventional Minkowski 
vacuum but instead a quantum state adapted to the expansion of the plasma. 

Such nonzero vacuum expectation values 
are commonplace in curved spacetime quantum field theory, 
and they do occur also in flat spacetime, 
in particular for vacua that 
are adapted to special families of 
(possibly noninertial) observers~\cite{byd}. 
An example that is relevant for us is the Rindler vacuum. 
This is a state defined in the quadrant 
$x>|t|$ of Minkowski space and seen as a 
no-particle state by the family of the 
observers given by 
\begin{subequations}
\ba
t &=& \xi \sinh\eta , 
\\
x &=& \xi \cosh\eta , 
\\
x^i  &=& b^i, \ \ i = 2,3, 
\ea
\end{subequations}
where the constants $b^i \in \mathbb{R}$ 
and $\xi>0$ specify the observer's trajectory
and $\eta$ equals $1/\xi$ times the observer's proper time. 
Each trajectory follows an orbit of a boost in $(t,x)$ and 
has uniform linear acceleration 
of magnitude $1/\xi$~\cite{letaw}. 
For a conformal scalar field, the energy-momentum 
in the Rindler vacuum has a nonvanishing expectation value~\cite{takagi}: 
using $(\eta,\xi,x^2,x^3)$ as the coordinates, the metric reads 
\be
ds^2_{\mathrm{Rindler}}=-\xi^2 d\eta^2 + d\xi^2 
+{(dx^2)}^2+{(dx^3)}^2 , 
\label{eq:rindler}
\ee
and the energy-momentum tensor is given in these coordinates by 
\be
T_\mu{}^\nu
= \frac{1}{1440 \pi^2 \xi^4}
\,\,
\mathrm{diag}
\left(
3, -1, -1, -1
\right) .
\label{T-rindler}
\ee

Now, the Milne-like coordinates $(\tau,\rapidity,x^2,x^3)$ are defined in the quadrant $t> |x|$ of Minkowski space, and they are adapted to the Bjorken flow in the sense that the velocity vector of the flow is~$\partial_\tau$. 
The metric in these coordinates reads 
\be
ds^2_{\mathrm{Milne}}=-d\tau^2+\tau^2d\rapidity^2 
+{(dx^2)}^2+{(dx^3)}^2 . 
\label{eq:milne}
\ee
The Rindler energy-momentum tensor \nr{T-rindler} is singular at $t=x$, but if it is analytically continued across this singularity into the quadrant $t> |x|$ and expressed in the coordinates $(\tau,\rapidity,x^2,x^3)$, it becomes
\be
T_\mu{}^\nu
= \frac{1}{1440 \pi^2 \tau^4}
\,\,
\mathrm{diag}
\left(
-1, 3, -1, -1
\right) . 
\label{T-r-cont}
\ee
This energy-momentum tensor 
has thus exactly the $\tau$-dependence of the 
last term displayed in~\nr{epsconfexp}. 
Could the last term displayed in \nr{epsconfexp} therefore be 
a vacuum energy contribution that should be subtracted before 
reading off from \nr{epsconfexp} the energy density due 
to excitations of the fluid? Note that this term is independent of~$T_f$, 
and it is the only term in \nr{epsconfexp} that survives in the limit $T_f\to0$. 

A case in which such a Casimir-type vacuum energy term is present, 
and indeed crucial for obtaining consistent scale-free thermodynamics, is 
the $(1+1)$-dimensional Bjorken flow~\cite{klt}. 
The Casimir term in the holographic energy-momentum tensor 
is identified from the limit of a vanishing bulk black hole and is given by 
\be
\epsilon
=p
= - \frac{\CL}{16 \pi G_3 \tau^2} , 
\label{eq:1+1-Casimir}
\ee
where $\CL$ is the length scale 
of the bulk cosmological constant. 
This Casimir term 
duly has the form 
of the energy-momentum tensor of a conformal scalar field in the 
appropriate conformal 
vacuum adapted to the expanding fluid flow~\cite{klt}. 
There is also evidence that a similar Casimir term could be present in
the spatially isotropic counterpart of the Bjorken flow in $d\ge3$
dimensions, with $\epsilon$ and $p$ proportional to
$1/\tau^d$~\cite{klt3}.

In this paper we attempt to identify the prospective Casimir
contribution to the boundary energy-momentum tensor by assuming that
the corresponding bulk solution has more symmetry, by one more Killing
vector, than what the symmetries of the boundary Bjorken flow require.
We find the bulk solution explicitly, and we show that it is locally
just an unusual foliation of the Schwarzschild-AdS${}_5$ ``bubble of
nothing''
\cite{Aharony:2002cx,Birmingham:2002st,Balasubramanian:2002am}.  The
holographic energy-momentum tensor turns out to have the
form~\nr{T-r-cont}, with an overall 
coefficient that is proportional to the
mass parameter of the bulk solution. This holographic energy-momentum
tensor is a limiting case of the family of boost-invariant
energy-momentum tensors considered in~\cite{jp}, and our bulk solution
can thus be considered as completing part of the programme
initiated in~\cite{jp}. 
However, we show that 
the bulk solution has always a singularity, either curvature or
conical, except when the solution reduces to pure AdS${}_5$ and 
the holographic energy-momentum tensor vanishes. 
Our bulk solution does therefore not provide compelling evidence for a
nonvanishing Casimir energy-momentum tensor. 

We begin by briefly reviewing in Section \ref{sec:B-ansatz} the
Bjorken flow ansatz on the boundary and the corresponding metric
ansatz in the bulk. In Section 
\ref{sec:bulksol} we specialise the bulk 
ansatz in a way that gives the bulk an additional Killing vector, 
solve the field equations and 
find the holographic energy-momentum tensor. 
The global properties of the solution are 
discussed in Section~\ref{sec:global}, 
and the possible interpretation of the 
holographic energy-momentum tensor 
in terms of quantum 
field theory is discussed 
in Section~\ref{sec:Tmn}. 
Section \ref{sec:conclusions} presents a summary and 
discusses the prospects of identifying a 
Casimir term under weaker assumptions.

\section{Bjorken flow and its dual ansatz}
\label{sec:B-ansatz}

Recall that in Milne-like coordinates
$(\tau,\rapidity,x^2,x^3)$ 
Minkowski metric takes the form~\nr{eq:milne}, 
and these coordinates are adapted to the Bjorken flow so that the
velocity vector of the flow is~$\partial_\tau$. The boost-invariance
and the transverse translational invariance of the flow imply that the
hydrodynamic variables are independent of the rapidity $\rapidity$ 
and the transverse spatial coordinates $(x^2,x^3)$. 
Assuming the matter to be a conformally invariant perfect
fluid, whose energy-momentum tensor satisfies $T_\mu^\mu=0$ and
$\nabla_\mu T^\mn=0$, and working 
in the coordinates~\nr{eq:milne}, 
it can be shown that
\be
T^\mu_{\,\,\,\nu}=\left( \begin{array}{cccc}
  -\epsilon(\tau)& 0 & 0&0 \\
  0 & -\epsilon(\tau)-\tau \epsilon'(\tau) & 0 &0\\
  0 & 0 & \epsilon(\tau)+\fr12\tau \epsilon'(\tau) &0\\
  0 & 0 & 0 & \epsilon(\tau)+\fr12\tau \epsilon'(\tau)
  \end{array} \right), 
  \label{tmunugen}
\ee
where the only undetermined function is the energy density
$\epsilon(\tau)=T_{\tau\tau}=-T^\tau_{\,\,\,\tau}$. 
If the energy-momentum tensor satisfies the weak energy condition,
$T_\mn t^\mu t^\nu\ge0$ for any timelike vector $t^\mu$,
$\epsilon(\tau)$ must satisfy 
\cite{jp}
\be
\epsilon(\tau)\ge0,\qquad -4\epsilon(\tau)\le\tau\epsilon'(\tau)\le0.
\label{limits}
\ee
In the special case in which 
$\epsilon(\tau)$ has the power-law behaviour $\tau^{-p}$, 
the energy-momentum tensor takes the form 
\be
T^\mu_{\,\,\,\nu}=\epsilon(\tau)\left( \begin{array}{cccc}
  -1& 0 & 0&0 \\
  0 & p-1 & 0 &0\\
  0 & 0 & 1-\half p&0\\
  0 & 0 & 0 & 1-\half p
  \end{array} \right),  
  \label{tmunup}
\ee
and the weak energy condition \nr{limits} 
then implies $0\le p\le 4$.

By the symmetries of the Bjorken flow, one may expect its 
five-dimensional gravity dual to be of the form \cite{jp}
\be
ds^2 =  {\CL^2 \over z^2} \left\{ - a(\tau,z) d\tau^2
  + \tau^2 b(\tau,z)d\rapidity^2
  +c(\tau,z) \left[ {(dx^2)}^2+{(dx^3)}^2 \right] + dz^2   \right\},
  \label{jpmetric}
\ee
where the positive functions 
$a(\tau,z)$, 
$b(\tau,z)$ and 
$c(\tau,z)$ 
are such that the metric
satisfies five-dimensional Einstein's equations 
with the cosmological constant~$-6/\CL^2$, 
\be
R_{MN}=-{4\over\CL^2}g_{MN}, 
\label{adseq}
\ee
and 
$a(\tau,z)$, 
$b(\tau,z)$ and 
$c(\tau,z)$ all 
tend to 1 as $z \to 0$.  
Once the dual solution is found, its 
holographic energy-momentum tensor can be computed 
from the asymptotic small $z$ expansion of the metric, 
\begin{subequations}
\label{eq:fg-exp}
\ba
ds^2 &=& {\CL^2\over z^2} [ g_\mn dx^\mu dx^\nu+dz^2 ], 
\label{AdS-form-in-z}
\\
g_\mn(\tau,z) &=& g^{(0)}_\mn(\tau)
+g^{(2)}_\mn(\tau)z^2+g^{(4)}_\mn(\tau)z^4+ \ldots ,
\label{gexp}
\ea
\end{subequations} 
where $g^{(0)}_\mn$ is the Milne
metric~\nr{eq:milne}: the result is 
\cite{skenderis}
\begin{align}
T_\mn=\frac{\CL^3}{4\pi G_5}\left[g^{(4)}_{\mu\nu}
-\fra18 g^{(0)}_\mn[(\Tr g_{(2)})^2-\Tr
   (g_{(2)}^2)]
   - \fra12 (g_{(2)}g_{(0)}^{-1}g_{(2)})_\mn
   + \fra14 (\Tr g_{(2)}) \,
   g_{(2)\mn} \right] .
\label{eq:Tmunu-d=4}
\end{align}

\section{Bulk ansatz with an additional isometry}
\label{sec:bulksol}

We look for a bulk solution in which the functions $a(\tau,z)$,
$b(\tau,z)$ and $c(\tau,z)$ in the ansatz 
\nr{jpmetric} depend on $\tau$ and $z$ solely through
the combination $s \equiv {(z/\tau)}^2$. Geometrically, this means
that in addition to the Killing vector $\partial_\rapidity$ and the three
Killing vectors that generate the $E_2$ isometries in $(x^2,x^3)$, the
metric admits also the Killing vector $\tau\partial_\tau + z\partial_z
+ x^2 \partial_{x^2} + x^3 \partial_{x^3}$.

Writing $a(s) =  s h^2(s)$, where $h>0$ and $s>0$, Einstein's equations yield
for $h$ the single ordinary differential equation
\be
h \left[ 
h^2 - 4 s^2 {(h')}^2\right] 
= (h^2-1) 
\left(
4 s^2 h'' + 4sh' - h \right) , 
\label{eq:hmaster}
\ee
where the prime denotes derivative with respect to~$s$. Once $h$ is
found, the functions $b(s)$ and $c(s)$ are given in terms of quadratures as 
\ba
\log\frac{b(s)}{b(0)} &=& 
\int_0^s 
ds \, 
\frac{{(sh^2)}' 
\left[ 3h^2 - 1 - {(sh^2)}' \right]}
{s^2(h^2-1) {(h^2)}'} , 
\label{eq:blogint}
\\
\log\frac{c(s)}{c(0)} &=& 
\int_0^s 
ds \, 
\frac{{(sh^2)}'}{s(h^2-1)} . 
\label{eq:clogint}
\ea

To solve~\nr{eq:hmaster}, 
we write it in terms of $\log s$ as the independent variable. 
This makes the equation autonomous, and it can integrated by
regarding $dh/d(\log s)$ as a function of~$h$. With the boundary
condition $a(s) = s h^2(s) \to1$ as $s\to0$, we find that $h(s)$ is
determined implicitly by
\be
\frac{1}{\sqrt{s}} = 
h(s) \exp\left\{
\int_{h(s)}^\infty dh 
\left[ 
\frac{1}{h}
- \frac{\sqrt{h^2-1}}{\sqrt{h^2(h^2-1) - \mu}}
\right]
\right\} , 
\label{eq:hintres}
\ee
where $\mu$ is a dimensionless constant of integration. 
We take $\mu$ to be real-valued. 
Equations 
\nr{eq:blogint} and \nr{eq:clogint} 
and the definition of $h$ then yield 
\begin{subequations}
\label{eq:abc-intres}
\ba
a&=& s h^2 , 
\label{eq:aintres}
\\
b&=& 
\frac{s 
\left[ 
h^2 ( h^2 -1 ) - \mu
\right]}
{h^2 -1} , 
\label{eq:bintres}
\\
c &=& 
s (h^2-1) 
\exp 
\left[
2 \int_h^\infty 
\frac{dh}{\sqrt{h^2-1} 
\sqrt{h^2(h^2-1) - \mu}}
\right] , 
\label{eq:cintres}
\ea
\end{subequations} 
where we have adopted the boundary conditions $b \to1$ and $c \to1$ as
$s\to0$ and the argument $s$ is being suppressed. 
Note that the
integrals in
\nr{eq:hintres} and \nr{eq:cintres} converge at infinity. 

The functions $h(s)$, $a(s)$, 
$b(s)$ and $c(s)$ are well defined for sufficiently 
small~$s$, and the small $s$ expansions 
of the metric coefficients read 
\begin{subequations}
\label{eq:abc-small}
\ba
a(s) &=& 1 + {\textstyle \frac14} \mu s^2 + O(s^3) , 
\\
b(s) &=& 1 - {\textstyle \frac34} \mu s^2 + O(s^3) , 
\\
c(s) &=& 1 + {\textstyle \frac14} \mu s^2 + O(s^3) . 
\ea
\end{subequations} 
From 
\nr{eq:fg-exp}, 
\nr{eq:Tmunu-d=4}
and \nr{eq:abc-small} 
we then find that the holographic energy-momentum tensor
in the coordinates of \nr{eq:milne}
is given by 
\be
T_\mu{}^\nu
= \frac{\CL^{3}}{4 \pi G_{5}}
\frac{\mu}{4\tau^4}
\,\,
\mathrm{diag}
\left(
1, -3, 1, 1
\right) .
\label{Tmn}
\ee
Note the similarity with~\nr{T-r-cont}.

\section{Global properties of the bulk solution} 
\label{sec:global} 

To analyse the global properties of the bulk solution
given by \nr{eq:hintres}
and~\nr{eq:abc-intres}, 
we first 
replace the coordinates 
$(\tau,z)$ by $(\gamma, h)$, where $h$ is as in \nr{eq:hintres} and 
\be
\tau = \CL 
\exp
\left[
\gamma + 
\int_h^\infty 
\frac{dh}{\sqrt{h^2-1} 
\sqrt{h^2(h^2-1) - \mu}}
\right] . 
\ee
We then write 
$h = \sqrt{1+{(\rho/\CL)}^2}$, where $\rho>0$. 
For given~$\mu$, these transformations are well defined
for sufficiently small $s$, and the corresponding regime in the
coordinates $(\gamma, \rapidity, x^2, x^3, \rho)$ is that of
sufficiently large~$\rho$. The metric takes the form 
\ba
ds^2 &=& 
\left( 
\frac{\rho^2}{\CL^2} + 1 - \frac{\mu\CL^2}{\rho^2}
\right) \CL^2 d\rapidity^2 
% \nonumber
% \\
% &&
+ \, 
\frac{d\rho^2}
{ \displaystyle \left( 
\frac{\rho^2}{\CL^2} + 1 - \frac{\mu\CL^2}{\rho^2}
\right)}
\nonumber
\\
&&
+ 
\rho^2 
\left\{
- d\gamma^2 + e^{-2\gamma} 
\CL^{-2} \left[{(dx^2)}^2 + {(dx^3)}^2 \right]
\right\} . 
\label{eq:bubblemetric}
\ea

The metric \nr{eq:bubblemetric} is recognised as a double analytic
continuation of Schwarz\-schild-AdS${}_5$~\cite{gibbons}. 
$\rho$ is the usual Schwarzschild radial coordinate, 
$\CL\rapidity$ is the
Euclidean Schwarzschild time, and the part multiplied by 
$\rho^2$ is the metric on $(2+1)$-dimensional
de~Sitter space written in the spatially flat coordinate patch
$(\gamma,x^2, x^3)$, 
\be
ds^2_{dS_3} = 
- d\gamma^2 + e^{-2\gamma} 
\CL^{-2} \left[{(dx^2)}^2 + {(dx^3)}^2 \right] , 
\ee
which is the analytic continuation of the round unit $S^3$ to
Lorentzian signature. 
The
parameter $\mu$ is proportional to the Schwarzschild
mass. The isometry group is seven-dimensional, consisting of 
translations in $\rapidity$ and the six-dimensional isometry group
of $(2+1)$ de~Sitter space. 
Einstein's equations have therefore resulted into two more
Killing vectors than the four that we assumed in the metric ansatz.

The global properties of the solution depend on the sign of~$\mu$:
\begin{itemize}
\item  
When $\mu>0$, the metric \nr{eq:bubblemetric} is locally the 
Schwarzschild-AdS ``bubble of nothing''
\cite{Aharony:2002cx,Birmingham:2002st,Balasubramanian:2002am}. The
range of $\rho$ is $\rho_+ < \rho$, where
\be
\rho_+ = \CL \sqrt{\sqrt{\mu + 
{\textstyle\frac14}} - {\textstyle\frac12}} , 
\label{eq:rhoplus-def}
\ee
and there is a conical singularity at $\rho\to\rho_+$. 
If $\rapidity$ were periodic with period 
\be
% \frac{\pi \sqrt{\sqrt{\mu + 
% {\textstyle\frac14}} 
% - {\textstyle\frac12}}}{\sqrt{\mu + {\textstyle\frac14}}}
% = 
\frac{2\pi (\rho_+/\CL)}{2 {(\rho_+/\CL)}^2 + 1} , 
\label{eq:bubble-rapidity-period}
\ee
the conical singularity would be replaced by a bolt-type 
\cite{Gibbons:1979xm}
fixed point
set of the Killing vector~$\partial_\rapidity$, 
and the metric
\nr{eq:bubblemetric} would then be the genuine 
``bubble of nothing''. Periodicity in 
$\rapidity$ does however not appear physically 
appropriate for modelling ion collisions on the boundary. 
\item 
When $\mu<0$, the metric \nr{eq:bubblemetric} has a scalar curvature
singularity at $\rho\to0$. 
\item 
When $\mu=0$, the metric \nr{eq:bubblemetric} is locally AdS${}_5$,
and $\rho\to0_+$ is a coordinate singularity on a null
hypersurface. Tracing back to the form of the metric in~\nr{jpmetric},
this solution reads 
$a(\tau,z) = b(\tau,z) = c(\tau,z) =1$, and the null hypersurface
$\rho=0$ is at $\tau=z$. 
\end{itemize}

To end this section, we recall that 
the usual way of attaching to the metric
\nr{eq:bubblemetric} a conformal boundary 
is via hypersurfaces of constant~$\rho$. 
Replacing $\rho$ by the coordinate $\zeta$ by 
\be 
\frac{\rho^2}{\CL^2}
= 
\frac{\CL^2}{\zeta^2}
- \frac12
+ \frac{(\mu+\frac14) \zeta^2}{4 \CL^2} , 
\ee
the metric takes the usual Fefferman-Graham form, 
\be
ds^2 = 
\frac{\CL^2}{\zeta^2}
\left\{
\left[1 
- \frac{\zeta^2}{2\CL^2}
+ \frac{(\mu+\frac14) \zeta^4}{4 \CL^4}
\right] \CL^2 ds^2_{dS_3}
% \nonumber
% \\
% &&
+ 
\frac{\left[ \displaystyle 1 - 
\frac{(\mu+\frac14) \zeta^4}{4 \CL^4}\right]^2}
{\left[\displaystyle 1 
- \frac{\zeta^2}{2\CL^2}
+ \frac{(\mu+\frac14) \zeta^4}{4 \CL^4}\right]}
 \CL^2 d\rapidity^2 
% \nonumber
% \\
% &&
+ d\zeta^2
\right\} . 
\ee
and the boundary metric at $\zeta\to0$ is 
\be
ds^2_{\mathrm{bubble-b}}
= 
\CL^2 \left(
d\rapidity^2 + ds^2_{dS_{3}}\right) . 
\label{eq:bubble-boundary}
\ee
In the coordinates of~\nr{eq:bubble-boundary}, the 
holographic energy-momentum tensor reads \cite{Balasubramanian:2002am} 
\be
T_\mu{}^\nu=
\frac{\CL^3}{4\pi G_5} \frac{\mu+\frac14}{4\CL^4}
\,\,
\mathrm{diag}
\left(
-3, 1, 1, 1
\right) .
\label{tmunustatic-4}
\ee
As discussed in the
context of the spatially isotropic Bjorken flow in~\cite{klt3}, 
the transformation between \nr{Tmn} and 
\nr{tmunustatic-4} is compatible with the 
four-dimensional conformal anomaly and with the 
conformal transformation between
the boundary metrics 
\nr{eq:milne}
and~\nr{eq:bubble-boundary}, 
\be
ds^2_{\mathrm{Milne}} = {(\tau/\CL)}^2 ds^2_{\mathrm{bubble-b}} 
\ee
with $\tau/\CL = e^\gamma$.

\section{Casimir interpretation of the holographic energy-momen\-tum tensor?} 
\label{sec:Tmn}

We wish to discuss whether the energy-momentum tensor~\nr{Tmn}, 
for some 
nonvanishing value of~$\mu$, could be present as a 
Casimir part in the holographic energy-momentum 
tensor computed from the less symmetric bulk solution in 
\cite{jp,jp2,nakamura1,nakamura2, janik1,janik2,%
kt,kovchegov,bbhj,buchel,Heller:2008mb,Kinoshita:2008dq}. 
As immediate consistency checks for such an interpretation, 
we note that $T_\mn$ \nr{Tmn} is traceless, 
and it is invariant
under the longitudinal boosts and under the $E_2$ isometries in
$(x^2,x^3)$. 
Also, $T_\mn$ is singular at the collision event and on its light cone. 
Finally, $T_\mn$ has the form of the energy-momentum tensor 
\nr{T-r-cont} that came by analytically continuing the Rindler vacuum energy-momentum tensor \nr{T-rindler} from the quadrant $x>|t|$ to the Bjorken flow region, with the overall sign agreeing if $\mu<0$. 

However, there is both a bulk argument and a boundary 
argument against such an interpretation. 

The bulk argument is that the bulk solution has a singularity for every nonvanishing~$\mu$, as discussed in section~\ref{sec:global}. 
Even if 
such a bulk singularity is regarded as physically acceptable from the boundary viewpoint, the bulk geometry seems not to provide a criterion for fixing a distinguished nonzero value of~$\mu$.  

As a preparation for the boundary argument, recall \cite{klt} 
that in the case of the 
$(1+1)$-dimensional boundary, 
the Casimir term 
\nr{eq:1+1-Casimir}
has the form 
of the energy-momentum tensor of a conformal scalar field in 
the conformal vacuum state, 
defined by the massless limit of the mode functions 
(5.38) of~\cite{byd}. If the mass in the mode functions 
(5.38) of \cite{byd} is strictly positive, on the other hand, 
it can be verified that 
the mode sum expression for the 
vacuum polarisation $\langle\phi^2 \rangle$ remains divergent at small spatial momentum even after 
the Minkowski vacuum contribution is subtracted mode by mode, using 
(5.41) of~\cite{byd}. 
This indicates that the state is not Hadamard 
and standard techniques do not furnish it with a well-defined energy-momentum tensor \cite{Kay:1988mu,radzikowski}. 
Physically, this state is pathological at small spatial momentum since 
the mode functions in 
(5.38) of \cite{byd} become in this limit their own complex conjugates. 

Now, consider a conformal scalar field on the 
$(3+1)$-dimensional boundary~\nr{eq:milne}. The 
symmetries suggest that the prospective vacuum 
state with the energy-momentum tensor 
\nr{Tmn} should be defined in terms of mode 
functions whose dependence on $(\tau,\rapidity)$ 
is as in (5.38) of~\cite{byd}, 
with the effective mass coming from the 
Fourier-momenta in $(x^2,x^3)$. 
However, the mode sum expression for
$\langle\phi^2 \rangle$ is again divergent 
even after mode-by-mode subtraction of the Minkowski 
vacuum contribution, indicating that the state is not 
Hadamard and does not have a well-defined energy-momentum tensor. 
This suggests that the energy-momentum tensor 
\nr{Tmn} may not be the vacuum energy-momentum tensor of any state that is regular in the Hadamard sense of \cite{Kay:1988mu,radzikowski}, even though it is related to the Rindler vacuum energy-momentum tensor by analytic continuation 
acrosss the Rindler horizon. 

Note that both of these objections disappear if $\mu>0$ and 
$\rapidity$ is periodic with the period~\nr{eq:bubble-rapidity-period}. 
In the bulk the periodicity removes the conical singularity. 
On the boundary periodicity of $\rapidity$ makes 
the $\rapidity$-momentum 
discrete, and the divergence in the mode sum integral for 
$\langle\phi^2 \rangle$ in the 
limit of small $\rapidity$-momentum is then no longer present. 
Periodicity in 
$\rapidity$ does however not appear physically 
appropriate in the ion collision setting, 
as we mentioned after~\nr{eq:bubble-rapidity-period}.

\section{Conclusions}
\label{sec:conclusions}

We have asked whether the holographic 
energy-momentum tensor found in the AdS/CFT 
description of heavy ion collisions 
in the Bjorken flow approximation 
should be attributed in its entirety to the expanding matter, 
or whether part of it should be interpreted as a Casimir-like 
vacuum energy-momentum term. 
This question is prompted by the observation that 
in the corresponding 
$(1+1)$-dimensional Bjorken flow problem such a Casimir term is 
present, and this term is indeed crucial for 
consistency of the scale-free hydrodynamic
approximation beyond the high density limit~\cite{klt}. 

We postulated a bulk ansatz that assumes one more Killing vector than 
those of the boundary Bjorken flow. We found the corresponding 
bulk solution in terms of quadratures, we showed that the holographic energy-momentum 
has the anticipated form, and we showed that this bulk solution is locally just the 
Schwarzschild-AdS${}_5$ ``bubble of nothing'' in an unusual foliation. 
However, this bulk solution 
has a singularity except when it reduces to pure AdS${}_5$. 
Our bulk solution does therefore not 
provide compelling evidence for a
nonvanishing Casimir part in the energy-momentum tensor of 
the approximate Bjorken flow bulk solutions 
analysed in 
\cite{jp,jp2,nakamura1,nakamura2,janik1,janik2,%
kt,kovchegov,bbhj,buchel,Heller:2008mb,Kinoshita:2008dq}. 

We show in the Appendix that our bulk solution can be readily generalised 
so that the scale factor of the transverse 
boundary dimensions equals ${(\tau/\CL)}^p$ 
with $-\infty < p \le 1$, with $p=0$ being the Bjorken flow case. 
For $0<p\le1$, the boundary metric is then an expanding cosmology, 
and one might attempt to interpret the holographic 
energy-momentum tensor as that of plasma in an expanding cosmology. 
A~problem with such an interpretation is however that the $p\ne0$ 
bulk solution has
the same singularities as the $p=0$ solution. 

To summarise, our $(4+1)$-dimensional bulk ansatz 
did not lead to a viable Casimir term in the 
$(3+1)$-dimensional boundary energy-momentum tensor. 
While our additional bulk 
Killing vector is an obvious generalisation of the 
Killing vector that does lead to 
a viable Casimir term 
in the lower-dimensional setting of a  
$(2+1)$-dimensional bulk and a $(1+1)$-dimensional boundary~\cite{klt}, 
might one perhaps have fared better 
by postulating a different Killing vector 
in the $(4+1)$-dimensional bulk ansatz?  
We shall now argue from the symmetries of the 
AdS solution that this is unlikely. 

Recall that in coordinates adapted to the $d$-dimensional 
boundary Bjorken flow, the 
pure AdS${}_{d+1}$ bulk solution reads  
\begin{subequations}
\ba
ds^2 &=& \frac{\CL^2}{z^2} 
\left[ -d\tau^2 + \tau^2 d\rapidity^2 + dz^2 \right] , 
\qquad \text{(for $d=2$)}
\label{eq:poinc-milne-2}
\\
ds^2 &=& \frac{\CL^2}{z^2} 
\left[ -d\tau^2 + \tau^2 d\rapidity^2 
+ {(dx^2)}^2 + \cdots + {(dx^{d-1})}^2 + dz^2 \right] , 
\qquad \text{(for $d\ge3$)}
\label{eq:poinc-milne-higher}
\ea
\end{subequations}
where 
$\rapidity$ is the longitudinal rapidity coordinate 
and the transverse coordinates $x^i$ are present only for $d\ge3$. 
This is the solution one would a priori expect to be the bulk ground state. 
Now, the $d=2$ solution \nr{eq:poinc-milne-2}
has the Killing vector 
$\tau\partial_\tau + z\partial_z$, which is timelike near the infinity and 
commutes
with the Bjorken flow Killing vector $\partial_\rapidity$. 
It is this Killing vector, via the temperature and entropy of its Killing horizon, 
that makes it possible to interpret the solution 
\nr{eq:poinc-milne-2} as giving the 
boundary Bjorken flow a nonzero temperature and entropy~\cite{klt}. 
In the $d\ge3$ solution~\nr{eq:poinc-milne-higher}, by contrast, 
the only Killing vectors that commute with both the Bjorken flow 
longitudinal Killing vector $\partial_\rapidity$ and 
the transversal Killing vectors $\partial_{x^i}$ 
can be verified to be 
linear combinations of these 
Killing vectors themselves.\footnote{We thank Don Marolf for raising this question.} 
The Killing vector of our ansatz, 
$\tau\partial_\tau + z\partial_z
+ \sum_i x^i \partial_{x^i}$, 
does not commute with the symmetries of the Bjorken flow, 
nor can it be replaced with one that would. 
The Killing horizon argument of $d=2$ does therefore not generalise 
into a thermodynamical Bjorken flow 
interpretation of the $d\ge3$ AdS solution~\nr{eq:poinc-milne-higher}. 

That being said, the $d=4$ 
bulk solution that is known in terms of its late time expansion 
is known to have an event horizon~\cite{Kinoshita:2008dq}. 
The possibility of a Casimir contribution in the 
holographic energy-momentum tensor should perhaps  
be examined in a formalism that allows genuinely 
time-dependent notions of temperature and entropy~\cite{ashtekar-dynamical}.

\vspace{1cm}
{\it Acknowledgements}.
We thank John Billingham for 
help with equation~\nr{eq:hmaster}, 
Chris Fewster, Don Marolf 
and Mitch Pfenning for discussions and correspondence 
and 
Makoto Natsuume
for bringing references to our attention. 
This research has been supported in part by Academy of Finland,
contract number 109720,
by the National Science Foundation under Grant No.\ HY05-51164, 
by STFC (UK) grant PP/D507358/1
and by NSERC, Canada. 
KK thanks the Institute for Nuclear Theory at the
University of Washington for its hospitality and
the Department of Energy for partial support
during the completion of this work.
JL thanks Helsinki Institute of Physics for
hospitality in the final stages of this work.

\begin{appendix}

\section{Appendix: Non-constant transverse dimensions} 

In this appendix we show how the bulk solution 
found in the main text generalises to give a boundary metric in which 
the scale factor of the transversal dimensions equals ${(\tau/\CL)}^p$ 
with $-\infty < p \le 1$. 

We start by generalising the bulk metric 
\nr{eq:bubblemetric} to 
\ba
ds^2 &=& 
\left( 
\frac{\rho^2}{\CL^2} + k - \frac{\mu\CL^2}{\rho^2}
\right) \CL^2 d\rapidity^2 
% \nonumber
% \\
% &&
+ \, 
\frac{d\rho^2}
{ \displaystyle \left( 
\frac{\rho^2}{\CL^2} + k - \frac{\mu\CL^2}{\rho^2}
\right)}
\nonumber
\\
&&
+ 
\rho^2 
\left\{
- d\gamma^2 + e^{-2\sqrt{k} \, \gamma} 
\CL^{-2} \left[{(dx^2)}^2 + {(dx^3)}^2 \right]
\right\} , 
\label{eq:bubblemetric-gen}
\ea
where $k$ is a non-negative constant. 
For $k>0$, the metric \nr{eq:bubblemetric-gen} is obtained from 
\nr{eq:bubblemetric} by first 
doing the coordinate transformation 
\be
(\rapidity, \rho, \gamma, x^2, x^3) 
= (\sqrt{k} \, \tilde \rapidity, 
{\tilde\rho}/\sqrt{k} , 
\sqrt{k} \, {\tilde\gamma} , 
\sqrt{k} \, {\tilde x^2} , 
\sqrt{k} \, {\tilde x^3})
\ee
with $\mu = {\tilde \mu}/k^2$, and then dropping the tildes. 
Taking the limit $k\to0$ in \nr{eq:bubblemetric-gen} 
does not bring in qualitative changes for our purposes: 
the $k=0$ metric \nr{eq:bubblemetric-gen} is still 
locally AdS${}_5$ when $\mu=0$, it has a conical singularity when 
$\mu>0$ and a scalar curvature singularity when $\mu<0$. 
Note that when $\mu>0$, the $k=0$ metric is locally the 
AdS${}_5$ soliton~\cite{Horowitz:1998ha}, 
and it would be globally the AdS${}_5$
soliton if $\rapidity$ were periodic with period~$\pi/\mu^{1/4}$. 

We now 
generalise the coordinate transformation of Section 
\ref{sec:global} to~\nr{eq:bubblemetric-gen}. 
Working at sufficiently large~$\rho$, 
we replace $\rho$ by 
$h = \sqrt{1+{(\rho/\CL)}^2}$ 
and define $s$ and $\tau$ by 
\ba
\frac{1}{\sqrt{s}} 
&=& 
h \exp\left\{
\int_{h}^\infty dh 
\left[ 
\frac{1}{h}
- \frac{\sqrt{h^2-1}}{\sqrt{{(h^2-1)}^2 + k (h^2-1) - \mu}}
\right]
\right\} , 
\label{eq:hintres-k}
\\
\tau &=& \CL 
\exp
\left[
\gamma + 
\int_h^\infty 
\frac{dh}{\sqrt{h^2-1} 
\sqrt{{(h^2-1)}^2 + k (h^2-1) - \mu}}
\right] . 
\ea
We then define $z = \tau \sqrt{s}$, or $s = {(z/\tau)}^2$, 
and write the metric in the 
coordinates $(\tau, \rapidity, x^2, x^3, z)$, 
where $h$ is defined as a function of $s$ by~\nr{eq:hintres-k}. 
We find that the metric takes the form \nr{jpmetric} with 

\begin{subequations}
\begin{align}
a &= s h^2 , 
\\
b &= 
\frac{s 
\left[ 
{(h^2-1)}^2 + k (h^2-1) - \mu
\right]}
{h^2 -1} , 
\label{eq:bintres-k}
\\
c &= 
{\left( \frac\tau\CL \right)}^{2(1-\sqrt{k})}
\, s (h^2-1) 
\exp 
\left[
2 \sqrt{k} \int_h^\infty 
\frac{dh}{\sqrt{h^2-1} 
\sqrt{{(h^2-1)}^2 + k (h^2-1) - \mu}}
\right] , 
\label{eq:cintres-k}
\end{align} 
\end{subequations} 
where all three expressions are functions 
of the suppressed argument~$s$, they are well defined 
for sufficiently small~$s$, and their small $s$ expansions read 
\begin{subequations}
\ba
a(s) &=& 1 + {\textstyle \frac12} (1-k) s 
+ \bigl[ 
{\textstyle \frac14} \mu + {\textstyle \frac1{16}} {(k-1)}^2
\bigr] s^2 + O(s^3) , 
\\
b(s) &=& 
1 + {\textstyle \frac12} (k-1) s 
+ \bigl[ 
{- \textstyle \frac34} \mu + {\textstyle \frac1{16}} {(k-1)}^2
\bigr] s^2 + O(s^3) , 
\\
c(s) &=& 
{\left( \frac\tau\CL \right)}^{2(1-\sqrt{k})}
\left\{
1 - {\textstyle \frac12} {(\sqrt{k} -1)}^2 s 
+ \left[ 
{\textstyle \frac14} \mu + {\textstyle \frac1{16}} {(\sqrt{k}-1)}^4
\right] s^2 + O(s^3)
\right\} . 
\ea
\end{subequations} 
The conformal boundary metric is therefore 
\be
ds^2=-d\tau^2+\tau^2d\rapidity^2 
+ {\left( \frac\tau\CL \right)}^{2(1-\sqrt{k})}
\left[ 
{(dx^2)}^2+{(dx^3)}^2
\right] . 
\label{eq:tau-cosmo}
\ee
From~\nr{eq:Tmunu-d=4}, the holographic energy-momentum tensor 
reads 
\begin{align}
T_\mu{}^\nu
&= \frac{\CL^{3}}{4 \pi G_{5}}
\frac{1}{4\tau^4}
\,\,
\Bigl\{
\mu \, \mathrm{diag}
\left(
1, -3, 1, 1
\right) 
\Bigr. 
\nonumber
\\[1ex]
&
\hspace{6ex}
\Bigl. 
+ \fra14 \, \mathrm{diag}
\left( 
{(\sqrt{k} -1)}^3 {(\sqrt{k} +3)} ,  
- {(\sqrt{k} -1)}^3 {(3\sqrt{k} +1)} ,  
{({k} -1)}^2 , 
{({k} -1)}^2 
\right)
 \Bigr\} , 
\label{Tmn-k}
\end{align} 
reducing to \nr{Tmn} for $k=1$. 

In the boundary metric~\nr{eq:tau-cosmo}, the scale factor of the
transverse dimensions equals ${(\tau/\CL)}^p$, where $p=
1-\sqrt{k}$. Note that $-\infty < p \le 1$. 
The Ricci scalar of the metric equals $R = 4p^2/\tau^2$: 
for $p\ne0$, the metric hence does 
not satisfy Einstein's vacuum equations 
and has a scalar curvature singularity at $\tau\to0$. 
For $0 < p \le1$, the metric can thus be understood
as an expanding cosmology, with an initial singularity at $\tau=0$, 
and the expansion is isotropic when $p=1$. In the range $0 < p \le1$, 
or $0 \le k <1$, 
one might therefore attempt to interpret the energy-momentum tensor 
\nr{Tmn-k} as that of plasma in an expanding
cosmology. 
However, as the bulk singularities 
are qualitatively similar for all~$k$, 
the objections that were discussed in the main text 
for $k=1$ are present also for other values of~$k$.

\end{appendix}

\end{document}